\documentstyle[aps,multicol,epsfig]{revtex}
\def\be{\begin{eqnarray}}
\def\ee{\end{eqnarray}}

\renewcommand{\narrowtext}{\begin{multicols}{2} \global\columnwidth20.5pc}
\renewcommand{\widetext}{\end{multicols} \global\columnwidth42.5pc}

\def\inseps#1#2{\def\epsfsize##1##2{#2##1} \centerline{\epsfbox{#1}}}

\def\top#1{\vskip #1\begin{picture}(290,80)(80,500)\thinlines \put(
65,500){\line( 1, 0){255}}\put(320,500){\line( 0, 1){
5}}\end{picture}}
\def\bottom#1{\vskip #1\begin{picture}(290,80)(80,500)\thinlines \put(
330,500){\line( 1, 0){255}}\put(330,500){\line( 0, -1){
5}}\end{picture}}

\begin{document}
\title{Holon Pair Condensation and Phase Diagram of
High $T_c$ Cuprates}
\author{T.-H.~Gimm$^\ast$,
S.-S.~Lee$^\dagger$,
 S.-P.~Hong$^\ddagger$,
 and S.-H.~Suck Salk$^\S$}
\address{Department of Physics,
Pohang University of Science and
Technology\\ Pohang 790-784, Korea}
\date{December 7, 1998}

\maketitle

\begin{abstract}
A possibility of 
holon (boson) pair condensation is explored 
for hole doped high $T_c$ cuprates, by using
the U(1) slave-boson representation 
of the $t$-$J$ Hamiltonian with the inclusion of hole-hole repulsion.
A phase diagram of the hole doped high $T_c$ cuprates is deduced
by allowing both the holon pairing and spinon pairing.
It is shown that the spin gap size remains nearly unchanged
below the holon pair condensation temperature.
We find that the $s$-wave holon pairing 
under the condition of $d$-wave singlet pairing is preferred, thus
allowing $d$-wave hole pairing.
\end{abstract}

\bigskip

\def\be{\begin{eqnarray}}
\def\ee{\end{eqnarray}}
\narrowtext
\section{Introduction}
It is believed that the high $T_c$ superconductivity
arises as a result of Bose condensation of doped holes.
It is still an unresolved problem
to predict a satisfactory phase diagram of
involving 
the bose condensation of the doped holes and
the normal state pseudogap.
The normal state pseudogap has
been observed in various experiments
\cite{nmr,ishida,neutron,caxis,uchida,loram,bucher,oda,loeser,ding}:
NMR \cite{nmr,ishida}, neutron scattering \cite{neutron}, $c$-axis
optical conductivity \cite{caxis,uchida}, heat capacity \cite{loram}, 
in-plane resistivity \cite{bucher,oda} and photoemission \cite{loeser,ding}.
A universal dependence of superconducting
critical temperature $T_c$ on hole doping rate 
is observed
for various high $T_c$ cuprates, 
by manifesting
a smooth increase of $T_c$ up to
the optimal doping rate and a decrease of $T_c$ beyond it.
This observation is well fit by a universal relation,
$T_c/T_c^{\rm max} = 1 - 82.6(x-0.16)^2$
where $T_c^{\rm max}$ is
the maximum critical temperature
at the optimal doping rate of $x = 0.16$
\cite{presland}.

Based on the $t$-$J$ model Hamiltonian, the 
normal state pseudogap is regarded
as the spin gap \cite{anderson,kotliar,fukuyama,ioffe,lee,ubbens,ubbens2}.
Earlier superconductivity in the underdoped
cuprates was understood as a simultaneous presence of
the spin gap and single boson (holon) condensation 
\cite{kotliar,fukuyama,ioffe,lee,ubbens,ubbens2}.
Recently an SU(2) slave-boson theory of the $t$-$J$ model Hamiltonian
was proposed to allow symmetry at both half filling and
finite doping \cite{wen1,wen2}.
In this theory \cite{wen1} 
single boson (holon) condensation can be
either completely destroyed or reduced
due to a low lying fluctuation (soft mode) in association with
the SU(2) rotation,
and thus boson (holon) pair condensation is suggested \cite{wen2}.
Earlier various preformed-pair scenarios
\cite{doniac,emery,randeria} were proposed.
In one of the scenarios \cite{emery}, 
the preformed pairs become locally available
below the pseudogap temperature, and
the critical temperature of superconductivity
is determined by
phase ordering of the preformed pairs.
On the other hand, 
numerical calculations 
\cite{dagotto,scalapino,white,riera} have been made to study
the paring of doped holes 
in the $t$-$J$ clusters and $t$-$J$ ladder systems.
A recent study
for doped $t$-$J$ three-leg ladders \cite{rice}
revealed hole pairing due to the coupling of a 
Luttinger liquid to the insulating or doped spin liquid.
In view of great interest in the role of doped holes 
for high T$_c$ superconductivity,
we investigate a possibility of
boson pair condensation by introducing a modified $t$-$J$ Hamiltonian
and derive a phase diagram in the plane of temperature 
vs. hole doping rate.

\section{Mean Field Hamiltonian and Free Energy from
a Modified \lowercase{$t$}-$J$ Hamiltonian in the Slave-Boson
Representation}
A necessity of introducing the
electrostatic hole-hole repulsion
is stressed in recent numerical
studies of a spin ladder system \cite{riera,gazza}.
In order to account for a reasonable hole pair binding
a large value of 
Coulomb repulsion energy $V$
between two
nearest neighbor (NN)
holes was needed, i.e.,
$V=e^2/(\epsilon_{\infty}c_0)=0.1{\rm eV}$ \cite{riera}
with $c_0 \simeq 3.8{\rm \AA}$, the lattice constant and
$\epsilon_{\infty} = 30 \sim 40$,
the dielectric constant \cite{chen}.
Ignoring this repulsion energy, the bound state of the hole pairs was
found to be excessively robust \cite{gazza}.

In the local U(1) symmetry conserving slave-boson representation,
we write 
the full $t$-$J$ Hamiltonian
of the two dimensional systems of antiferromagnetically
correlated electrons by including a hole-hole
repulsion term,
\be \label{H}
H && =
-t\sum_{\langle i,j \rangle\sigma}
\left (f^{\dagger}_{i\sigma}b_ib^{\dagger}_jf_{j\sigma} + h.c.
 \right)  
+ J\sum_{\langle i,j \rangle}
\left ( {\bf S}_i\cdot {\bf S}_j - \frac{n_in_j}{4} \right )  \nonumber \\
&& +V\sum_{\langle i,j \rangle}
b^{\dagger}_ib_ib^{\dagger}_jb_j
-\mu_0\sum_if^{\dagger}_{i\sigma}f_{i\sigma}
\ee
with ${\bf S}_i =
 1/2f^{\dagger}_{i\sigma}{\bf \sigma}_{\alpha\beta}f_{i\beta}$.
Here the
local constraint of single occupancy, $\sum_\sigma f^{\dagger}_{i\sigma}f_{i\sigma} +
 b^{\dagger}_ib_i = 1$ is assumed.
$f^{\dagger}_{i\sigma} (f_{i\sigma})$ is the spinon
creation (annihilation) operator and
$b_i (b^{\dagger}_i)$, the holon annihilation (creation)
 operator.
The third term represents
Coulomb repulsion between holes in the NN sites; according to
an experiment \cite{tjeng} 
the hole-hole repulsion decays rapidly with distance.
The second bracketed term 
above implies that
the NN configuration of two
holes is energetically more favorable than 
other possible configurations. This is because
the two holes in the
NN sites break
7 bonds compared to 8 bonds for other configurations.
This is evident from the separate inspection of the two
attractive interaction
terms, $J {\bf S}_i\cdot {\bf S}_j$
and $-(J/4) n_in_j$. 
For the latter we write,
\be \label{ninj}
&&-J\sum_{\langle i,j \rangle}\frac{n_in_j}{4}
= -\frac{J}{4}\sum_{\langle i,j \rangle}
\left \{ 1 - b^{\dagger}_ib_i - b^{\dagger}_jb_j
+ b^{\dagger}_ib^{\dagger}_jb_ib_j \right \} \nonumber \\
&&= -\frac{J}{2}\sum_{i\sigma}f^{\dagger}_{i\sigma}f_{i\sigma}
+\frac{J}{2}\sum_ib^{\dagger}_ib_i
- \frac{J}{4}\sum_{\langle i,j \rangle}
b^{\dagger}_ib^{\dagger}_jb_ib_j \/.
\ee
where the local constraint of  single occupancy is considered.
The effective attraction between
the NN holes
arises 
from the last term of the equation above.

Applying the Hubbard-Stratonovich transformation \cite{ubbens} in Eq.~(\ref{H}),
both the Heisenberg term
and the hopping term
are converted into 
linearized terms involving the hopping order field
$\chi_{ji} = \langle {8t}/{3J}b^{\dagger}_jb_i 
+ f^{\dagger}_{j\sigma}f_{i\sigma} \rangle$ 
for the 
exchange interaction channel and 
the spinon pairing order field
$\Delta^{{f}}_{ji}
 = \langle f_{j\uparrow}f_{i\downarrow}-f_{j\downarrow}f_{i\uparrow}
\rangle $ 
for the pairing channel.
Such reduction to the two channels is made 
by ignoring the direct (Hartree) channel based on 
the assumption of
paramagnetic states for each site, i.e., $\langle {\bf S}_i \rangle = 0$ 
\cite{ubbens}. Thus 
long-range antiferromagnetic fluctuations 
are ignored.
The resulting effective Hamiltonian is then,
\widetext
\top{-2.8cm}
\be \label{H00}
&& H = \sum_{\langle i,j \rangle}\frac{3J}{8} \left [ |\chi_{ji}|^2
+ |\Delta_{ji}^{{f}}|^2
 - \left (\frac{8t}{3J}b^{\dagger}_jb_i +
f^{\dagger}_{j\sigma}f_{i\sigma}\right )\chi_{ji}
- h.c.
- (f_{j\uparrow}f_{i\downarrow}-f_{j\downarrow}f_{i\uparrow})
{\Delta_{ji}^{f}}^{\ast} - h.c.  \right ] \nonumber \\
&& + \frac{8t^2}{3J}\sum_{\langle i,j \rangle}(b^{\dagger}_j
b_i)(b^{\dagger}_ib_j)
 -\sum_{\langle i,j \rangle}
(\frac{J}{4}-V) b^{\dagger}_ib^{\dagger}_jb_ib_j 
-(\mu_0-\frac{1}{4})\sum_if^{\dagger}_{i\sigma}f_{i\sigma}
+\frac{J}{2}\sum_ib^{\dagger}_{i}b_{i} \\
&&-i\sum_i \lambda_i(f^{\dagger}_{i\sigma}f_{i\sigma}
+b^{\dagger}_ib_i - 1 ) \/, \nonumber
\ee
\bottom{-2.8cm}
\narrowtext
\noindent
where a Lagrange multiplier field $\lambda_i$ is
introduced for the local
constraint of single occupancy for both the spinon and the holon.
The quartic holon term (the second term in Eq.~(\ref{H00}) above),
$\frac{8t^2}{3J}\sum_{\langle i,j \rangle}(b^{\dagger}_j
b_i)(b^{\dagger}_ib_j)$ is repulsive \cite{baskaran}.
It is important to realize that this term corresponds to 
the holon exchange interaction, but not to 
the direct (forward)
 interaction nor to the holon pairing interaction.

Thus allowing only the holon exchange channel,
we linearize 
the quartic holon term (the second term in Eq.~(\ref{H00})) as
\widetext
\top{-2.8cm}
\be 
\label{key}
\frac{8t^2}{3J}\sum_{\langle i,j \rangle}(b^{\dagger}_j
b_i)(b^{\dagger}_ib_j) = 
\frac{8t^2}{3J}\sum_{\langle i,j \rangle}
\left (\langle b^{\dagger}_jb_i \rangle
b^{\dagger}_ib_j  +
b^{\dagger}_jb_i
\langle b^{\dagger}_ib_j \rangle - \langle b^{\dagger}_jb_i\rangle
\langle b^{\dagger}_ib_j \rangle \right )\/.
\ee
We decompose the 
effective holon attractive 
energy term (the third term in Eq.~(\ref{H00})), that is,
$-\sum_{\langle i,j \rangle}
(\frac{J}{4}-V) b^{\dagger}_ib^{\dagger}_jb_ib_j$ 
with $0 < V < J/4$ 
 into the direct,  exchange, and pairing channels \cite{hfb}.
By introducing the Hubbard-Stratonovich transformation for the
resulting holon pairing term and
the above linearized holon exchange term,
we obtain
\be \label{H0}
&& H  = \sum_{\langle i,j \rangle}\frac{3J}{8} \left [ |\chi_{ji}|^2
+ |\Delta_{ji}^{{f}}|^2
 - \left (\frac{8t}{3J}b^{\dagger}_jb_i +
f^{\dagger}_{j\sigma}f_{i\sigma}\right )\chi_{ji}
- h.c.
- (f_{j\uparrow}f_{i\downarrow}-f_{j\downarrow}f_{i\uparrow})
{\Delta_{ji}^{f}}^{\ast} - h.c.  \right ] \nonumber \\
&& +\sum_{\langle i,j \rangle}
(\frac{J}{4}-V)
\left [
|\Delta^{b}_{ji}|^2
- b^\dagger_jb^\dagger_i{\Delta^{b}_{ji}}^\ast - h.c. \right ]
- \sum_{\langle i,j \rangle}
(\frac{8t^2}{3J} - \frac{J}{4} + V)
\left [ \langle b^{\dagger}_jb_i\rangle
\langle b^{\dagger}_ib_j \rangle
- b^{\dagger}_jb_i\langle b^{\dagger}_ib_j \rangle
 - h.c. \right ]  \\
&&-\mu^{f}_0\sum_if^{\dagger}_{i\sigma}f_{i\sigma}
-\mu^{b}_0\sum_ib^{\dagger}_{i}b_{i}
-i\sum_i \lambda_i(f^{\dagger}_{i\sigma}f_{i\sigma}
+b^{\dagger}_ib_i - 1 ) \/, \nonumber
\ee
\bottom{-2.8cm}
\narrowtext
\noindent
with $\mu^f_0 = \mu_0 -J/4$ and $\mu^b_0 = -3J/4$,
where $\Delta^{\rm b}_{ji}$ is
now the Hubbard-Stratonovich field for holon pairing.
Here $\mu^f_0$ and $\mu^b_0$ are
the effective chemical potentials for the spinon and the holon
respectively.
Obviously the two scalar fields,
$\chi_{ji} = \langle {8t}/{3J}b^{\dagger}_jb_i
+ f^{\dagger}_{j\sigma}f_{i\sigma} \rangle$
and $\langle b^{\dagger}_jb_i \rangle$ in Eq.~(\ref{H0}) above
are not independent.
We allow a linear relation between the spinon hopping and
the holon hopping order parameters, i.e.,
$\langle f^{\dagger}_{i\sigma}f_{j\sigma} \rangle =
\eta\langle b^{\dagger}_ib_j \rangle$. Thus we rewrite  
$\chi_{ji} = 
(8t/(3J)+2\eta)\langle b^{\dagger}_jb_i \rangle$ or 
\be \label{change}
\langle b^{\dagger}_jb_i \rangle  = 
\frac{1}{8t/(3J)+2\eta} \chi_{ji} \/.
\ee
A self-consistent determination of the hopping ratio $\eta$ 
will be discussed later.

Substitution of Eq.~(\ref{change}) into Eq.~(\ref{H0})
results in the cancelation of terms (proportional to $\frac{8t^2}{3J}$)
involving the holon exchange channel in Eq.~(\ref{H0}).
By defining
the pairing order parameters
$\Delta^{f}_{r} = \langle f_{j\uparrow}f_{j+{r}, \downarrow}
-f_{j\downarrow}f_{j+{r}, \uparrow} \rangle $ and
$\Delta^{b}_{r} = \langle b^\dagger_jb^\dagger_{j+{r}} \rangle$
with $r = \hat{x}$ or $\hat{y}$, 
and allowing the uniform hopping order parameter, $\chi_{ji} = 
\chi$,
the resulting mean field Hamiltonian is 
\be \label{Ha}
&& H_{\rm MF} = 
\sum_{\langle i,j \rangle}\left [{\tilde J} |\chi|^2
 - \left (\tilde{t} b^{\dagger}_jb_i +
\frac{3J}{8}f^{\dagger}_{j\sigma}f_{i\sigma}\right )\chi
- h.c. \right ] \nonumber \\
&& + \sum_{j,r={\hat{x} \hspace{0.1cm}{\mbox {\footnotesize \rm or}}
\hspace{0.1cm} \hat{y}}}
\frac{3J}{8} \left [ |\Delta^{f}_{r}|^2
- (f_{j\uparrow}f_{j+r,\downarrow}-f_{j\downarrow}f_{j+r,\uparrow})
{\Delta^{f}_{r}}^{\ast} - h.c.  \right ] \nonumber \\
&& +\sum_{j,r={\hat{x} \hspace{0.1cm} {\mbox {\footnotesize \rm or}}
\hspace{0.1cm}\hat{y}}}
(\frac{J}{4}-V)
\left [
|\Delta^{b}_{r}|^2
- b^\dagger_jb^\dagger_{j+r}{\Delta^{b}_{r}}^\ast - h.c. \right ]  \\
&&-\mu^{f}_0\sum_if^{\dagger}_{i\sigma}f_{i\sigma}
-\mu^{b}_0\sum_ib^{\dagger}_{i}b_{i}
-i\lambda \sum_i (f^{\dagger}_{i\sigma}f_{i\sigma}
+b^{\dagger}_ib_i - 1 ) \/, \nonumber
\ee
where 
\be \label{J} \tilde{J} = \frac{3J\eta^2/2 + 4\eta t + J/4
 -V}{(2\eta + 8t/(3J))^2}
\ee 
 and 
\be  \label{tau}
\tilde{t} = \frac{2\eta t+J/4-V}{2\eta + 8t/(3J)} \/.
\ee
From the hopping ratio, $\eta =
\frac{\langle f^{\dagger}_{i\uparrow}f_{j\uparrow} \rangle}
{\langle b^{\dagger}_ib_j \rangle} \propto \frac{1-x}{x}$ 
where $x$ is the hole doping rate, we note that
in the half-filling limit, i.e.,
$x\rightarrow 0$,
$\tilde{J}$ in Eq.~(\ref{J}) approaches $3J/8$.
As a result this satisfies the SU(2) 
symmetry \cite{affleck} of the spinon pairing order field and 
the spinon hopping order field,
that is, $|\Delta^f_{r}| = \chi$ at $x=0$.
Following a similar consideration
to that of Lee and Nagaosa \cite{lee}, we obtain 
the effective spinon mass, ${m_{f}}^{-1} = c_o^2\frac{3J}{4}\chi$
($c_0$ is the lattice constant) 
and the effective holon mass $m_b$, 
\be \label{mb}
{m_{b}}^{-1} = c_0^22\tilde{t} \chi =
 c_0^2\frac{2\eta t+J/4-V}{\eta + 4t/(3J)}\chi \/.
\ee
from the use of Eq.~(\ref{Ha}) with Eq.~(\ref{tau}).

By allowing
$d$-wave pairing for spinons ($\Delta^f_x = - \Delta^f_y \equiv
\Delta^f$) and 
$s$-wave ($\Delta^b_x = \Delta^b_y \equiv \Delta^b$) 
or $d$-wave pairing 
($\Delta^b_x = -\Delta^b_y \equiv \Delta^b$)
for holons,
we obtain the Fourier transformed
mean field Hamiltonian from Eq.~(\ref{Ha}),
\be
&& H_{\rm MF} =
2N\left [ \tilde{J}|\chi|^2 +
\frac{3J}{8}
|\Delta^{f}|^2
+(\frac{J}{4}-V)|\Delta^{b}|^2 \right ] 
  + N\lambda \nonumber \\ 
&& + \sum_{k\sigma}(\epsilon^{f}_k - \mu^{f})
f^{\dagger}_{k\sigma}f_{k\sigma}  
- \sum_k \Delta^{f}_k
(f^\dagger_{k\uparrow}f^\dagger_{-k\downarrow} +h.c.)   \\
&& + \sum_{k}(\epsilon^{b}_k - \mu^{b})b^{\dagger}_{k}b_{k} 
- \sum_{k}\Delta^{b}_k(b^{\dagger}_{k}b^\dagger_{-k}
+ h.c.) \nonumber \/.
\ee
Here
$\epsilon^{f}_k = -\frac{3J}{4}\chi \gamma_k$,
$\epsilon^{b}_k = -{2}\tilde{t}\chi \gamma_k$,
$\mu^{f} \equiv \mu_0^{f} + \lambda$, and $\mu^{b} \equiv 
\mu_0^{b} + \lambda$ 
with $\gamma_k \equiv \cos{k_x} + \cos{k_y}$.
$\Delta^{f}_k = \frac{3J}{4} \Delta^{f} \varphi_k$ 
for $d$-wave spinon pairing
with $\varphi_k \equiv \cos{k_x} - \cos{k_y}$,
$\Delta^{b}_k = 2(\frac{J}{4}-V)\Delta^{b}\gamma_k$ for
$s$-wave holon pairing and 
$\Delta^{b}_k = 2(\frac{J}{4}-V)\Delta^{b}\varphi_k$ for
$d$-wave holon pairing. 
$N$ is the total number of lattice sites.
From the Bogoliubov
transformation for both the fermion (spinon) and boson (holon)
operators, we obtain
the quasi-particle excitation energies,
$E^{{f}}_k = \sqrt{(\epsilon^{f}_k - \mu^{f})^2
 + {\Delta_k^{{f}}}^2}$
for spinons and
$E^{{b}}_k =
\sqrt{(\epsilon^{b}_k - \mu^{b})^2 -
{\Delta_k^{b}}^2}$
for holons.
The spinon pairing gap (or spin gap) 
is $E_{\rm g}^f(k) = |\Delta_k^f|
= |\frac{3J}{4} \Delta^{f} \varphi_k|$.
We note that 
the upper limit of $|{\Delta_k^{b}}|$ is $|\epsilon^{b}_k - \mu^{b}|$.
For both the $s$-wave holon pairing and $d$-wave holon pairing,
the maximum ($E^b_{\rm max}$) of the holon quasi-particle excitation energy
$E_k^b$ occurs at ${\bf k} = (\pi,\pi)$, 
whereas the minimum ($E^b_{\rm min}$) of 
$E_k^b$ occurs at ${\bf k} = (0,0)$
(for the $s$-wave holon pairing
$E^b_{\rm max/min} =
\sqrt{(4\tilde{t}\chi \mp \mu^{b})^2 -
[{(J-4V)\Delta^{b}}]^2}$ and
$E^b_{\rm max/min}= |4\tilde{t}\chi \mp \mu^{b}|$ (where $\mu^b < 0$)
for the $d$-wave holon pairing
($-$ for max. and $+$ for min.).

The mean field free energy at doping rate $x$ is obtained to be,
\be \label{free}
&& F_{\rm MF}(\chi, \Delta^{{f}},
\Delta^{{b}})/N = 2\tilde{J}|\chi|^2
+ \frac{3J}{4}{\Delta^{f}}^2 
+ (\frac{J}{2}-2{V}){\Delta^{b}}^2 \nonumber  \\
&& - 2T\sum_{k}\ln\left[\cosh(\beta E^{{f}}_k/2)
\right ] 
+ T\sum_{k}\ln\left[\sinh(\beta E^{{b}}_k/2)
 \right ]   \nonumber \\
&& + (\frac{1}{2}+x)\mu^{b} - x\mu^{f} \/. 
\ee
The minimization of the mean field free energy 
 with respect to the scalar fields,
$\chi, \Delta^{f},$ and $\Delta^{b}$ will be
numerically made
as a function of temperature $T$ and doping rate
$x$ \cite{size}.
From the computed mean field values of the order parameters
($\chi, \Delta^f$ and $\Delta^b$) and
the chemical potentials ($\mu^f$ and $\mu^b$),
we determine the hopping ratio
$\eta$ self-consistently with the use 
of $\sum_\sigma\langle f^{\dagger}_{i\sigma}f_{j\sigma} \rangle 
= -\frac{1}{2N}\sum_k\left [\gamma_k {(\epsilon^f_k
-\mu^{f})}/{E_k^{f}} \right ]\tanh(\beta E_k^{f}/2)$ and
$\langle b^{\dagger}_ib_j \rangle = 
\frac{1}{4N}\sum_k\left [ \gamma_k {(\epsilon^b_k
-\mu^{b})}/{E_k^{b}} \right ] \coth(\beta E_k^{f}/2)$.

\section{Computed Results of Boson (Holon) Pair Condensation and 
Spin Gap Temperatures; Phase Diagram 
of $T$-\lowercase{$x$} Plane}
In Fig.~1 
we show as a function of doping rate the predicted
spin gap 
($E^f_{\rm g}(\pi, 0) = \frac{3J}{4}\Delta^f$),
the rescaled hopping order parameter \cite{kotliar}
 ($\chi$ multiplied by $\frac{3J}{4}$)
 and  the holon pairing order parameter
$\Delta^{b}$ respectively at
various selected temperatures.
The self-consistently determined result of $\eta$ is
found to be $\eta \simeq 0.2(1-x)/x$ with $x > 0.01$ for $J=0.4t$.
For the present study,
$J=0.4t$ \cite{ubbens2} and $V = 0.24999J$
were chosen to best fit the experimental phase diagram
\cite{ding,presland,williams} in the plane of $T$ vs. $x$.
The predicted spin gap $E^f_{\rm g}(\pi, 0)$ 
($y$-axis on the left hand side) at ${\bf k} = (\pi, 0)$
decreases with the hole doping rate $x$ at all
temperatures.
We find that the minimum of the free energy occurs
only with the $s$-wave holon pairing, but not with
the $d$-wave holon pairing.
The $d$-wave holon pairing 
is found to be unstable.
In Fig.~1 the $s$-wave holon pairing order parameter $\Delta^{b}$
($y$-axis on the right hand side) is displayed
at each selected finite temperature.
It is seen to increase with $x$ at all temperatures.
We now examine 
\begin{figure}[b]
\inseps{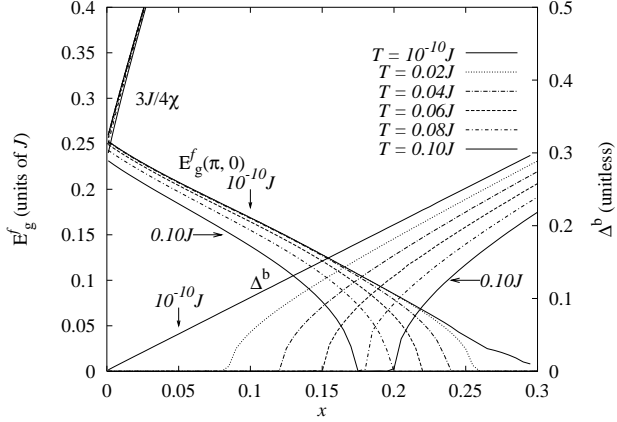}{0.66}
\caption{\label{fig1}
The computed
spin gap $E^f_{\rm g}$ (on the left axis)
at ${\bf k} = (\pi, 0)$,
the hopping order parameter
$\chi$ multiplied by $3J/4$
(on the left axis) and  holon pairing order parameter
$\Delta^{b}$ (on the right axis) at
various selected temperatures.}
\end{figure}
\noindent
the variation of the spin gap with
temperature below and above
the holon pair condensation temperature, $T^b_{\rm MF}$
in the underdoped region.
As an example, let us choose the doping rate of 
$x=0.1$, at which $T^b_{\rm MF}$ is obtained to be $0.028J$.
Two of the three nearly indistinguishable
lines of the spin gap $E^f_{\rm g}(\pi ,0)$
at $T=10^{-10}J$ and $T=0.02J$ 
(both of which 
represent temperatures below $T^b_{\rm MF}$)
indicate nearly identical spin gap sizes.
Thus the spin gap size
remains nearly unchanged below the holon pair condensation temperature.
The remaining four lines which correspond to temperatures above
$T^b_{\rm MF}$
show a consistent decrease in the spin gap 
size as temperature increases.

In Fig.~2, we present the computed
phase diagram in the $T$ -- $x$ plane.
The holon (boson) pair condensation temperature $T^{b}_{\rm MF}$
and the spin gap
temperature $T^{f}_{\rm MF}$ 
are plotted
as a function of doping rate $x$, including 
the spin gap $E^f_{\rm g}(\pi, 0)$ at $T = T^{b}_{\rm MF}$.
The mean field 
holon pair condensation temperature
is found to be less than $0.1J$ in the underdoped region. 
In this region both the $d$-wave spinon pairing 
with $\langle f^{\dagger}_{i\uparrow}f^{\dagger}_{j\downarrow}
\rangle \neq 0$
and the $s$-wave holon
pairing with $\langle b_ib_j \rangle \neq 0$
coexist below the mean field 
(critical) temperature ($T_c = T^b_{\rm MF}$).
For the hole (but not the holon) pairs,
we have $\langle c_{i\uparrow}c_{j\downarrow}  \rangle
= \langle b^\dagger_i b^\dagger_j \rangle \langle 
f_{i\uparrow} f_{j\downarrow} \rangle \neq 0$ in
the mean field approximation.
For clarity we would like to stress that the term,
`holon' refers to the boson of spin $0$ and the term, `hole' is
the fermion of spin up and spin down.
Thus this allows for the condensation of the
$d$-wave
hole pairs by satisfying the $s$-wave holon pairing for
$\langle b^\dagger_i b^\dagger_j \rangle$
and the
$d$-wave pairing for $\langle
f_{i\uparrow} f_{j\downarrow} \rangle$ 
in $\langle c_{i\uparrow}c_{j\downarrow} \rangle =
\langle b^\dagger_i b^\dagger_j \rangle \langle
f_{i\uparrow} f_{j\downarrow} \rangle$.
Thus 
$T^b_{\rm MF}$ at each doping rate 
corresponds to the mean field critical temperature
of the $d$-wave superconducting phase transition.

The mean field spin gap (pseudogap) temperature
$T^f_{\rm MF}$ may be regarded as 
the `center' of
crossover region which arises as a result of 
gauge fluctuations \cite{lee,ubbens2}. 
The predicted pseudogap (spin gap) temperature
$T^{f}_{\rm MF}$ is seen to
\begin{figure}[b]
\inseps{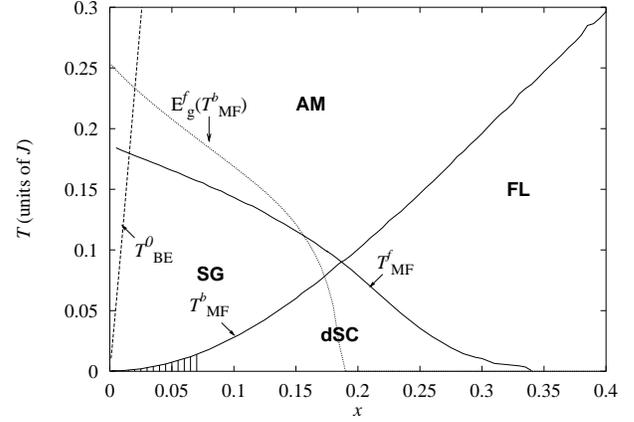}{0.66}
\caption{\label{fig2}
The phase diagram of holon pairing
and spinon pairing.
The AM (anomalous metallic), SG (spin-gap), FL (Fermi liquid), and
dSC ($d$-wave superconducting) phase regions are exhibited with
the holon-pair condensation temperature $T^{b}_{\rm MF}$ and
the $d$-wave spinon pairing temperature
$T^{f}_{\rm MF}$.
The spin gap $E^f_{\rm g}$ with ${\bf k} = (\pi, 0)$
at $T = T^{b}_{\rm MF}$ and the mean field
single holon condensation temperature denoted as
$T_{\rm BE}^{0}$ are also displayed.
The normal state gap $E^f_{\rm g}$ disappears
at $T^{f}_{\rm MF} = T^{b}_{\rm MF}$ with $x_{\rm cr} \sim 0.19$.
The shaded region represents the region at which
the holon pair condensation is unstable.}
\end{figure}
\vskip-0.2cm
\noindent
smoothly decrease with the hole doping rate as shown in Fig.~2. This
is consistent with the recent experiments
\cite{ishida,oda,ding}.
The spin gap $E^f_{\rm g}(\pi, 0)$ at
the $s$-wave holon pair
($d$-wave hole pair)
condensation temperature $T^{b}_{\rm MF}$ 
is predicted to show
a rapid decrease with $x$ as is shown in Fig.~2.
This spin gap (pseudogap)
vanishes
at the critical doping rate of $x_{\rm cr} \simeq 0.19$, at which
$T^{b}_{\rm MF}$ approaches $T^{f}_{\rm MF}$.
We find that the position of the critical doping rate 
is sensitive to the choice of $V$.
Unlike $T^b_{\rm MF}$,
$T^f_{\rm MF}$
is nearly independent of $V$.
The critical doping rate $x_{\rm cr}$ 
decreases as $V$ gets smaller.
With the choice of a vanishingly small value of $V$,
we find that the value of holon pair
order parameter $\Delta^b$ (and thus $T^b_{\rm MF}$) is 
excessively large even
at a small doping rate. 
This 
is consistent with a recent numerical study \cite{gazza}.
The choice of 
$V = 0.24999J$ fits best the experimental
value of the observed critical doping rate of 
$x_{\rm cr} = 0.19$ \cite{ding,williams}. 
For this case the effective NN attractive energy
is about $V_{\rm eff} = \frac{J}{4}-V \simeq 10^{-5}J$.
Earlier the mean field single boson condensation temperature 
was reported  to be
$T_{\rm BE}^{0} \simeq 2\pi x {c_0}^{-2}/m_b$ 
\cite{fisher,lee}. 
It is displayed in Fig.~2 for comparison with
our computed boson pair condensation temperature ($T_c = T^b_{\rm MF}$).
Obviously we find a wide difference between the two bose
condensation temperatures
of the holon-pair boson and the single holon boson.
The observed optimal doping rate of $x_{\rm op}^{\rm Exp} = 0.16$
\cite{presland} is known to
be smaller than the critical doping rate of $0.19$  \cite{ding,williams}.
In disagreement with observations,
our predicted boson pair 
(not single boson) condensation temperature
above the critical doping rate of $x_{\rm cr} = 0.19$
becomes larger than
the pseudogap (spin gap) temperature.
Earlier it was
suggested that the
\begin{figure}[b]
\inseps{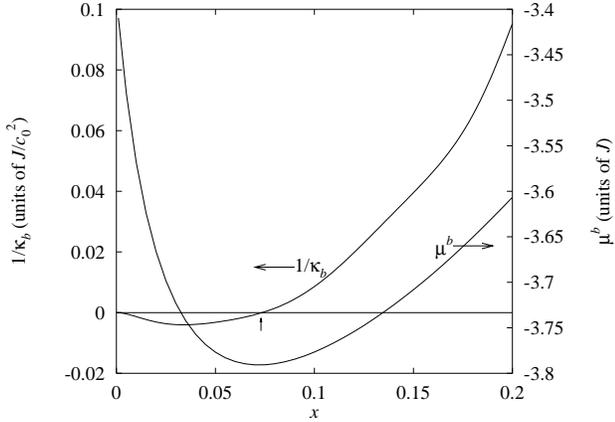}{0.66}
\caption{\label{fig3}
The effective holon
chemical potential $\mu^b(T =  0)$ (on the right axis)
and the inverse of the holon compressibility $1/\kappa_b$ (on the left axis)
as a function of doping rate $x$.}
\end{figure}
\noindent
overdoped region may correspond to the region
where the mean-field solution is not stable with respect to
the gauge fluctuations \cite{nagaosa} and that the system can probably be
described by the Fermi liquid theory \cite{lee,grilli}.
However, it remains to be seen for a
rigorous verification
in the future.

In Fig.~3 we present the effective holon
chemical potential $\mu^b$ at $T=0$ ($y$ axis on the right hand side) 
and the inverse of the holon compressibility $1/{\kappa_b}$ 
($y$ axis on the left hand side)
as a function of doping rate $x$.
The computed spinon chemical potential decreases monotonically with doping
rate $x$.
On the other hand,
the holon chemical potential decreases up to $x \simeq 0.07$ and
starts to increase beyond this value.
Allowing that 
the inverse of 
two dimensional holon
compressibility $\kappa_b$ holds for ${1}/{\kappa_b} =
\frac{{N}^2}{A}\frac{\partial \mu^b}{\partial N} =
\frac{x^2}{{c_0}^2}\frac{\partial \mu^b}{\partial x}$ \cite{nozieres}
(where $N$ is the number of bosons and $A$ is the area),
we find that ${1}/{\kappa_b}$
is negative below $x\sim 0.07$ and becomes positive for $x > 0.07$.
This suggests that in the region of
$x < 0.07$ the holon pair condensation is unstable \cite{nozieres}
and it is stable only for $x > 0.07$.
The shaded region in Fig.~2 represents the region at which
the holon pair condensation is unstable.
It is well known that the superconductivity
arises for $x > 0.05$ \cite{presland}.
Interestingly recent numerical studies\cite{white,rice} of
$t$-$J$ three-leg ladders revealed
that there exist a critical doping rate
of $x \sim 0.06$ beyond which 
the hole pairing increases rapidly as hole density increases.
It is of great interest to see in the future whether
there exists any 
possible relevance between the present result with 
the prediction of the $t$-$J$ three-leg ladders study.

\section{Conclusion}
A possibility of boson (holon)
pair condensation was described
with the inclusion of a repulsive interaction between the NN holes
in the $t$-$J$ Hamiltonian.
A phase diagram of the hole-doped high $T_c$ cuprates is 
derived by considering both the holon pair condensation and the
spinon pairing gap (spin gap).
In the mean field approximation, we note
that $\langle c_{i\uparrow}c_{j\downarrow}  \rangle
= \langle b^\dagger_i b^\dagger_j \rangle \langle
f_{i\uparrow} f_{j\downarrow} \rangle$ for the hole pair. 
To avoid confusion, we would like to point out
that the term, `hole' stands for the fermion of 
positive charge $+e$ with spin $1/2$, while the term,
`holon' refers to the boson of positive charge $+e$ with spin $0$.
We find that
the $s$-wave holon pairing ($\langle b^\dagger_i b^\dagger_j \rangle$)
but not the $d$-wave holon pairing is stable
for the $d$-wave spinon pairing
($\langle
f_{i\uparrow} f_{j\downarrow} \rangle$), 
thus  allowing 
the condensation of $d$-wave hole pairs in the language of 
`hole'.
In addition, it
is shown that the spin gap remains nearly unchanged below 
the boson pair condensation temperature.

One of us (SHSS) acknowledges the financial supports of Korea
Ministry of Education (BSRI-97) and the Center for 
Molecular Science at KAIST (1998). \\
\\
\noindent
$^\ast$thgim@anyon.postech.ac.kr \\
$^\dagger$ssong@anyon.postech.ac.kr \\
$^\ddagger$hong@anyon.postech.ac.kr \\
$^\S$salk@postech.ac.kr

\widetext

\begin{references}
\bibitem{nmr} 
Y.~Itoh, T.~Machi, S.~Adachi,
A.~Fukuoka, K.~Tanabe, H.~Yasuoka, J.~Phy.~Soc.~Jpn. {\bf 67}, 312 (1998);
references therein.
\bibitem{ishida}
K.~Ishida, K.~Yoshida, T.~Mito, Y.~Tokunaga, Y.~Kitaoka, K.~Asayama,
Y.~Nakayama, J.~Shimoyama, and K.~Kishio, Phys.~Rev.~B {\bf 58}, 5960 (1998).
\bibitem{neutron} J.~M.~Tranquada,
P.~M.~Gehring, and G.~Shirane, Phys.~Rev.~B {\bf 46}, 5561 (1992);
references therein.
\bibitem{caxis} C.~C.~Homes, T.~Timusk, R.~Liang, D.~A.~Bonn, and
W.~N.~Hardy, Phys.~Rev.~Lett. {\bf 71}, 1645 (1993).
\bibitem{uchida} S.~Uchida, K.~Tamasaku, K.~Takenaka, and Y.~Fukuzumi,
J.~Low~Temp.~Phys. {\bf 105}, 723 (1996).
\bibitem{loram} J.~W.~Loram, K.~A.~Mirza, J.~M.~Wade, 
J.~R.~Cooper, and W.~Y.~Liang,
Physica C {\bf 235-240}, 134 (1994).
\bibitem{bucher} B.~Bucher, P.~Steiner, J.~Karpinski,
E.~Kaldis, and P.~Wachter, Phys.~Rev.~Lett. {\bf 70}, 2012 (1993).
\bibitem{oda} M.~Oda, K.~Hoya, R.~Kubota, C.~Manabe, N.~Momono, T.~Nakano,
M.~Ido, Physica C {\bf 281}, 135 (1997).
\bibitem{loeser} A.~G.~Loeser, Z.~-X.~Shen, D.~S.~Dessau, D.~S.~Marshall,
C.~H.~Park, P.~Fournier, A.~Kapitulnik, Science {\bf 273}, 325 (1996).
\bibitem{ding} H.~Ding, 
T.~Yokoya, J.~C.~Campuzano, T.~Takahashi, M.~Randeria,
M.~R.~Norman, T.~Mochiku, K.~Kadowaki,
and J.~Giapintzakis, Nature {\bf 382}, 51 (1996);
H.~Ding, J.~C.~Campuzano, M.~R.~Norman, M.~Randeria,
T.~Yokoya, T.~Takahashi, T.~Takeuchi, T.~Mochiku, K.~Kadowaki,
P.~Guptasarma, and D.~G.~Hinks, J.~Phys.~Chem.~Solids in press;
cond-mat/9712100.
\bibitem{presland} M.~R.~Presland, J.~L.~Tallon, R.~G.~Buckley, R.~S.~Liu, and
N.~E.~Flower, Physica C {\bf 176}, 95 (1991); J.~L.~Tallon, 
C.~Bernhard, H.~Shaked, R.~L.~Hitterman, and J.~D.~Jorgensen,
Phys.~Rev.~B {\bf 51}, 12911 (1995).
\bibitem{anderson} 
G.~Baskaran, Z.~Zou, and P.~W.~Anderson, Solid State Commun. {\bf
63}, 973 (1987); G.~Baskaran and P.~W.~Anderson, Phys.~Rev.~B {\bf
37}, 580 (1988).
\bibitem{kotliar} G.~Kotliar and J.~Liu, Phys.~Rev.~B {\bf 38}, 5142 (1988).
\bibitem{fukuyama} Y.~Suzumura, Y.~Hasegawa, and H.~Fukuyama,
J.~Phys.~Soc.~Jpn. {\bf 57}, 401 (1988); H.~Fukuyama, 
Prog.~Theo.~Phys.~Suppl. {\bf 108}, 287 (1992).
\bibitem{ioffe} L.~B.~Ioffe and A.~I.~Larkin, Phys.~Rev.~B {\bf 39}, 8988
(1989).
\bibitem{lee} P.~A.~Lee and N.~Nagaosa, Phys.~Rev.~B {\bf 46}, 5621 (1992).
\bibitem{ubbens} M.~U.~Ubbens and P.~A.~Lee, Phys.~Rev.~B {\bf 46}, 8434 
(1992).
\bibitem{ubbens2} M.~U.~Ubbens and P.~A.~Lee,
 {\bf 49}, 6853 (1994); P.~A.~Lee, J.~Low~Temp.~Phys. {\bf 105}, 581 (1996).
\bibitem{wen1} X.-G.~Wen and P.~A.~Lee, Phys.~Rev.~Lett. {\bf 76},
503 (1996); P.~A.~Lee, N.~Nagaosa, T.-K.~Ng. X.-G.~Wen, 
Phys.~Rev.~B {\bf 57}, 6003 (1998).
\bibitem{wen2} X.-G.~Wen and P.~A.~Lee, Phys.~Rev.~Lett. {\bf 80}, 
2193 (1998); references therein.
\bibitem{doniac} S.~Doniac and M.~Inui, Phys.~Rev.~B {\bf 41}, 6668 (1990).
\bibitem{emery} V.~Emery and S.~Kivelson, Nature {\bf 374}, 434 (1995);
V.~J.~Emery, S.~A.~Kivelson, and O.~Zachar,
Phys.~Rev.~B {\bf 56}, 6120 (1997).
\bibitem{randeria} M.~Randeria, N.~Trivedi, A.~Moreo, and
R.~T.~Scalettar, Phys.~Rev.~Lett. {\bf 69},
2001 (1992).
\bibitem{dagotto} E.~Dagotto, A.~Nazarenko, and A.~Moreo, 
Phys.~Rev.~Lett. {\bf 74}, 310 (1995).
\bibitem{scalapino} S.~R.~White and D.~J.~Scalapino, Phys.~Rev.~B {\bf 55}, 6504
(1997).
\bibitem{white} S.~R.~White and D.~Scalapino, Phys.~Rev.~B {\bf 57},
 3031 (1998).
\bibitem{riera} J.~Riera and E.~Dagotto, Phys.~Rev.~B {\bf 57}, 8609
 (1998). 
\bibitem{rice} T.~M.~Rice, S.~Haas, and M.~Sigrist, F.~-C.~Zhang,
Phys.~Rev.~B {\bf 56}, 14655 (1997).
\bibitem{gazza} C.~Gazza, G.~B.~Martins, J.~Riera, and 
 E.~Dagotto, Phys.~Rev.~B in press; cond-mat/9803314.
\bibitem{chen} C.~Y.~Chen, E.~Ahrens, S.-W.~Cheong,
A.~Migliori, and Z.~Fisk, Phys.~Rev.~Lett. {\bf 63}, 2307 (1989);
D.~Reagor, N.~W.~Preyer, P.~J.~Picone, M.~A.~Kastner,
H.~P.~Jenssen, and D.~R.~Gabbe, A.~Cassanho, and
R.~J.~Birgeneau, {\it ibid}. {\bf 62}, 2048 (1989).
\bibitem{tjeng} L.~H.~Tjeng, H.~Eskes, and G.~A.~Sawatzky, in 
{\it Strong Correlation in Superconductivity}, Eds. H.~Fukuyama,
S.~Maekawa, and A.~P.~Malozemoff, Springer series in Solid State 
Science, Vol. 89 (1989).
\bibitem{baskaran} G.~Baskaran, Phys.~Scr.~T {\bf 27}, 53 (1989).
\bibitem{hfb} 
The decomposition of 
the four-boson term $-\frac{1}{4} b^{\dagger}_ib_ib^{\dagger}_jb_j$
into three different channels is made as follows:
\be
-\frac{1}{4} b^{\dagger}_ib_ib^{\dagger}_jb_j =
\hat{v}_{\rm E} + \hat{v}_{\rm P} + \hat{v}_{\rm D}  \nonumber
\ee
where
$\hat{v}_{\rm E}, 
\hat{v}_{\rm P}$, and  $\hat{v}_{\rm D}$ are the exchange,
pairing, and direct terms respectively.
Each term is explicitly obtained 
by satisfying the 
condition that a combination of any two terms among
the three terms above will not generate any
remaining term.
Using this condition 
they are, in the environment of antiferromagnetically
correlated electrons, 
\be \label{a7}
&& \hat{v}_{\rm E} =  -\frac{1}{4}
(b_{i}^\dagger b_{j})(b_{j}^\dagger b_{i})  \nonumber \\
&& \hat{v}_{\rm P} =  -\frac{1}{4}
(b_{i}^\dagger b_{j}^\dagger)(b_{j} b_{i}) \nonumber \\
&& \hat{v}_{\rm D} = +\frac{1}{4}
(b_{i}^\dagger b_{i})(b_{j}^\dagger b_{j})
= \frac{1}{4}n_in_j \/, \nonumber
\ee
where $n_i= b_{i}^\dagger b_{i}$ is holon number operator at site $i$.
\bibitem{affleck} I.~Affleck, Z.~Zou, T.~Hsu, and P.~W.~Anderson,
Phys.~Rev.~B {\bf 38}, 745 (1988).
\bibitem{size} The mean field free energy in Eq.~(\ref{free})
is calculated by 
discrete momentum summation rather than  the direct evaluation
of momentum space integral, in order 
to avoid singularities.
Such discretization allows us to determine the converged value of
the square lattice size.
A good convergence was found to occur with 
the square lattice of
$400 \times 400$ ($N  = 1.6 \times 10^5$).
\bibitem{williams} G.~V.~M.~Williams, J.~L.~Tallon, R.~Michalak, and
R.~Dupree, Phys.~Rev.~Lett. {\bf 78}, 721 (1997).
\bibitem{fisher} D.~S.~Fisher and P.~C.~Hohenberg, Phys.~Rev.~B {\bf 37}, 4936
 (1988).
\bibitem{nagaosa} N.~Nagaosa, Phys.~Rev.~Lett. {\bf 71}, 4210 (1993);
H.~Suzuura and N.~Nagaosa, Phys.~Rev.~B {\bf 56}, 3548 (1997).
\bibitem{grilli} M.~Grilli and G.~Kotliar,
 Phys.~Rev.~Lett. {\bf 64}, 1170 (1990).
\bibitem{nozieres} P.~Nozi$\grave{\rm e}$res and D.~Saint James,
J.~Physique {\bf 43}, 1133 (1982).
\end{references}
\end{document}